\documentclass[12pt,a4paper]{article}

\usepackage{amsmath}
\usepackage{amsfonts}

\newcommand{\be}{\begin{equation}}
\newcommand{\ee}{\end{equation}}
\newcommand{\bea}{\begin{eqnarray}}
\newcommand{\eea}{\end{eqnarray}}

\def\cA{{\cal A}}                       %
\def\ri{{\mathrm{i}}}                   %
\def\bR{{\mathbb R}}                    %
\def\bC{{\mathbb C}}                    %
\def\1{{\mbox{\boldmath $1$}}}          %
\def\tr{\mathrm{tr\,}}                  %
\def\cY{\mathcal{Y}}                    %
\def\cO{\mathcal{O}}                    %
\def\cC{\mathcal{C}}                    %
\def\cH{\mathcal{H}}                    %
\def\red{\mathrm{red}}                  %
\def\ad{\mathrm{ad}}                    %
\def\diag{\mathrm{diag}}                %
\def\cM{\mathcal{M}}                    %
\begin{document}

\vspace*{0.5cm}
\begin{center}
{\Large \bf  An application of the reduction method to Sutherland type many-body systems}
\end{center}

\vspace{0.2cm}

\begin{center}
L. Feh\'er \\

\bigskip

Department of Theoretical Physics, University of Szeged\\
Tisza Lajos krt 84-86, H-6720 Szeged, Hungary, and \\
Department of Theoretical Physics, WIGNER RCP, RMKI\\
H-1525 Budapest, P.O.B.~49,  Hungary\\
e-mail: lfeher@physx.u-szeged.hu

\bigskip

\bigskip

\end{center}

\vspace{0.2cm}

\begin{abstract}
We study Hamiltonian reductions of
the free geodesic motion on a non-compact simple
Lie group
using as reduction group the direct product of a maximal
compact subgroup and the fixed point subgroup of an
arbitrary involution commuting with the Cartan involution.
In general, we describe the reduced system that arises upon restriction to a dense open submanifold
and interpret it as a spin Sutherland system.
This dense open part yields the full reduced  system
in important special examples without spin degrees of freedom,  which include the  $BC_n$
Sutherland system built on 3 arbitrary couplings for
$m<n$ positively charged and $(n-m)$ negatively charged particles moving on the half-line.
\end{abstract}

\newpage
\section{Introduction}

One of the most popular approaches to integrable classical mechanical systems
is to realize systems of interest as reductions of higher dimensional
``canonical free systems''. The point is that
the properties of the reduced systems can be understood in elegant geometric terms.
This approach was pioneered by Olshanetsky and Perelomov \cite{OPCim} and by Kazhdan, Kostant and Sternberg
\cite{KKS}
who interpreted the celebrated rational Calogero and hyperbolic/trigonometric
Sutherland systems as Hamiltonian reductions of free particles moving on Riemannian
symmetric spaces.  As reviewed in \cite{OPRep,Banff,Suth},
these integrable many-body systems possess important generalizations
based on arbitrary root systems and
elliptic  interaction potentials.  They also admit
relativistic deformations,
extensions by  ``spin'' degrees of freedom
and generalizations describing  interactions of
charged particles.
The Hamiltonian reduction approach to many of these systems
was successfully worked out in the past (see e.g.~\cite{OPRep,FK} and their references),
but in some cases its discovery still poses us interesting open problems.

In a recent joint work with V. Ayadi \cite{AF}, we enlarged the range
of the reduction method to cover the $BC_n$ Sutherland system of charged particles
defined by the following Hamiltonian:
\bea
&& H= \frac{1}{2} \sum_{j=1}^n p_j^2
- \sum_{1\leq j\leq  m<k \leq n} (\frac{\kappa^2}{\cosh^2(q_j - q_k)} +
\frac{\kappa^2}{\cosh^2(q_j + q_k)} )
\nonumber\\
&&\phantom{X}
+ \sum_{1\leq j < k \leq m} ( \frac{\kappa^2 }{\sinh^2(q_j - q_k)} +
\frac{\kappa^2 }{\sinh^2(q_j + q_k)} )
+ \sum_{m<j< k\leq n} (\frac{\kappa^2}{\sinh^2(q_j - q_k)} + \frac{\kappa^2}{\sinh^2(q_j + q_k)} )
\nonumber\\
&&\phantom{XXX}
+ \frac{1}{2}\sum_{j=1}^n \frac{(x_0 - y_0)^2}{\sinh^2(2 q_j)}
+ \frac{1}{2}\sum_{j=1}^m \frac{x_0y_0}{\sinh^2(q_j)} -
\frac{1}{2}\sum_{j=m+1}^n \frac{  x_0y_0 }{\cosh^2( q_j)}.
\label{1}\eea
Here $m$ and $n$ are positive integers subject to $m<n$, while
$\kappa$, $x_0$ and $y_0$ are real coupling parameters satisfying
the conditions $\kappa\neq 0$ and   $(x_0^2 - y_0^2)\neq 0$, which permit
to consistently restrict the dynamics
to the domain where
\be
q_1>q_2>\cdots>q_m>0\qquad\hbox{and}\qquad
q_{m+1}> q_{m+2} >\cdots> q_n>0.
\label{2}\ee
If $x_0y_0>0$, then we can interpret
the Hamiltonian (\ref{1}) in terms of attractive-repulsive
 interactions between $m$ positively charged and $(n-m)$ negatively charged
particles influenced also by their mirror images and a positive charge
fixed at the origin.

The derivation \cite{AF} of
the Hamiltonian (\ref{1})  relied on reducing the free geodesic motion
on the group $Y:= SU(n,n)$ using  as symmetry group
$Y_+ \times Y^+$, where $Y_+<Y$ is a maximal compact subgroup and $Y^+<Y$ is
the (non-compact)
 fixed point subgroup of an involution of $Y$ that commutes with the
Cartan involution fixing $Y_+$.
This allowed us to cover the case of 3 arbitrary couplings, extending
the previous derivation \cite{Ha} of 2-parameter special cases of the system.
The $m=0$ special case was treated in \cite{FP} by applying the symmetry group $Y_+ \times Y_+$.

The emergence of
system (\ref{1}) as reduced system required to impose very special
 constraints on the free motion.
Thus it is natural to enquire about
the reduced systems that would arise under other moment map constraints.
In fact, the main purpose of this contribution is to characterize
the reduced systems in a general case, where $Y$
will be taken to be an arbitrary non-compact simple
Lie group, $Y_+ \times Y^+$ will have similar structure as mentioned
above, and the moment map constraint will be chosen arbitrarily.

In Section 2, we study reductions of the geodesic system on $Y$
restricting all considerations  to a dense open submanifold
consisting of regular elements.
In general, we shall interpret the reduced system as a  spin
Sutherland type system.
In exceptional cases, the initial restriction
to regular elements is immaterial in the sense that the moment map
constraint enforces the same restriction.
This happens in the reduction that yields the spinless system (\ref{1}),
as will be sketched in Section 3.
Finally, we shall present a short conclusion in Section 4.

\section{Spin Sutherland type systems from reduction}

We need to fix notations and recall an important group theoretic result
before turning to the reduction of our interest.

\subsection{Generalized Cartan decomposition}

Let $Y$ be a \emph{non-compact} connected simple real Lie group with Lie algebra $\cY$.
Equip $\cY$  with the scalar product $\langle\ ,\ \rangle$ given by a positive multiple
of the Killing form.
Suppose that $\Theta$ is a Cartan involution  of $Y$
(whose fixed point set is a maximal
compact subgroup) and $\Gamma$
is an arbitrary involution commuting with $\Theta$.
The corresponding involutions of $\cY$, denoted by $\theta$ and $\gamma$,
lead to the orthogonal decomposition
\be
\cY=\cY_+^+ + \cY_+^-+\cY_-^+ + \cY^-_-,
\label{3}\ee
where the subscripts $\pm$ refer to eigenvalues $\pm 1$  of $\theta$ and the superscripts to the
eigenvalues of
$\gamma$.
We may also use the associated projection operators
\be
\pi^\pm_\pm: \cY \to \cY^\pm_\pm,
\label{4}\ee
as well as $\pi_+= \pi_+^+ + \pi_+^-$ and $\pi^+=\pi^+_+ + \pi^+_-$.
We choose a maximal Abelian subspace
$$
\cA \subset \cY_-^-,
$$
and define
$$
\cC:= \operatorname{Cent}_{\cY}(\cA)= \{ \eta\in \cY\,\vert\,  [\eta,\alpha]=0 \,\,\,
\forall \alpha\in\cA\,\}.
$$
An element  $\alpha\in \cA$ is called \emph{regular} if its centralizer inside $\cY$
is precisely $\cC$.
The connected subgroup $A<Y$ associated with $\cA$
is diffeomorphic to $\cA$ by the exponential map.
For later use, we fix a connected component $\check \cA$ of the set of regular
elements of $\cA$, and introduce also the open submanifold
$$
\check A:= \exp(\check \cA)\subset A.
$$
The restriction of the scalar product to $\cC$ is non-degenerate
and thus we obtain the orthogonal decomposition
\be
\cY=\cC + \cC^\perp.
\label{8}\ee
According to (\ref{8}), any $X\in \cY$ can be written uniquely as $X=X_\cC + X_{\cC^\perp}$.
Equation (\ref{3}) induces also the  decomposition
$$
\cC= \cC_+^+ + \cC_+^- + \cC_-^+ + \cC^-_-,
\quad
\cC^-_-=\cA,
$$
and similarly for  $\cC^\perp$.

Let $Y_+$ and $Y^+$ be the fixed point subgroups of $\Theta$ and $\Gamma$, respectively,
possessing as their Lie algebras
$$
\cY_+ = \cY_+^+ + \cY_+^-
\quad\hbox{and}\quad
\cY^+=\cY^+_+ + \cY^+_-.
$$
Consider the group
$$
Y_+^+:= Y_+ \cap Y^+
$$
and its subgroup
\be
M:= \operatorname{Cent}_{Y^+_+}(\cA).
\label{12}\ee
Pretending that we deal only with matrix Lie groups,
the elements $m\in M$ have the defining property $m \alpha m^{-1} = \alpha$ for all
$\alpha \in \cA$. Note that $\cC_+^+$ is the Lie algebra of $M$.

We shall study the reductions of a free particle moving on $Y$
utilizing the symmetry group
$$
G:=Y_+ \times Y^+ < Y \times Y.
$$
It is a well-known group theoretic result (see e.g. \cite{KAH}) that every element $y\in Y$ can be written
in the form
\be
y= y_l a y_r
\quad\hbox{with}\quad
y_l\in Y_+,\, y_r\in Y^+,\, a\in A.
\label{14}\ee
This is symbolically expressed as the set-equality
\be
Y = Y_+ A Y^+.
\label{15}\ee
Furthermore, the subset of regular elements given by
\be
\check Y:= Y_+ \check A Y^+
\label{16}\ee
is open and dense in $Y$. The decomposition of $y\in \check Y$ in the form (\ref{14})
is unique up to the replacement $(y_l, y_r) \rightarrow (y_l m, m^{-1} y_r )$ with any $m\in M$.
The product decomposition (\ref{15}) is usually referred to as a generalized Cartan decomposition
since it reduces to the usual Cartan decomposition in the case $\gamma=\theta$.
This decomposition will play  crucial role in what follows.

\subsection{Generic Hamiltonian reduction}

We wish to reduce the Hamiltonian system of a free particle moving  on $Y$
along geodesics of the pseudo-Riemannian metric
associated with the scalar product $\langle\ ,\ \rangle$.
To begin, we trivialize $T^*Y$ by right-translations,  identify
$\cY$ with $\cY^*$ (and similarly for $\cY_+$ and $\cY^+$) by the scalar product,
and choose an
arbitrary coadjoint orbit
$$
\cO:= \cO^l \times \cO^r
$$
of the symmetry group $G= Y_+ \times Y^+$.
We then consider the phase space
$$
P:= T^*Y  \times \cO \simeq Y \times \cY \times \cO^l \times \cO^r = \{ (y,J, \xi^l, \xi^r)\}
$$
endowed with its natural symplectic form  $\omega$ and the free Hamiltonian $\cH$,
$$
\cH(y,J,\xi^l, \xi^r) := \frac{1}{2} \langle J, J\rangle.
$$
The form $\omega$ can be  written symbolically as $\omega = d \langle J, (dy) y^{-1}\rangle + \Omega$, where
$\Omega$ is the canonical symplectic form of the orbit $\cO$.

The action of $(g_l, g_r) \in G$ on $P$ is defined by
$$
\Psi_{(g_l, g_r)}: (y,J, \xi^l, \xi^r) \mapsto (g_l y g_r^{-1}, g_l J g_l^{-1}, g_l \xi^l g_l^{-1},
g_r \xi^r g_r^{-1}).
$$
This Hamiltonian action is generated by the moment map $\Phi= (\Phi_l, \Phi_r): P \to \cY_+ \times \cY^+$
whose components are
$$
\Phi_l(y,J,\xi^l, \xi^r) =\pi_+(J) + \xi^l,
\qquad
\Phi_r(y,J,\xi^l,\xi^r) = -\pi^+( y^{-1} J y) + \xi^r.
$$
We restrict our attention to the  ``big cell'' $\check P_\red$ of the full reduced phase space
\be
P_\red := \Phi^{-1}(0)/G
\label{22}\ee
that arises as  the symplectic reduction of the dense open submanifold
$$
\check P:= T^* \check Y \times \cO \subset P.
$$
In other words, we wish to describe the set of $G$-orbits,
\be
\check P_\red := \check P_c /G,
\label{24}\ee
in the constraint surface
\be
\check P_c:= \Phi^{-1}(0) \cap \check P.
\label{25}\ee

An auxiliary symplectic reduction of the orbit $(\cO,\Omega)$ by the group $M$ (\ref{12})
will appear in our final result. Notice that $M$ acts naturally on $\cO$ by its diagonal
embedding into $Y_+ \times Y^+$, i.e., by the symplectomorphisms
\be
\psi_m: (\xi^l, \xi^r) \mapsto (m \xi^l m^{-1}, m \xi^r m^{-1}),
\qquad
\forall m\in M.
\label{26}\ee
This action has its own moment map $\phi: \cO \to (\cC_+^+)^* \simeq \cC_+^+$ furnished  by
$$
\phi: (\xi^l, \xi^r) \mapsto \pi_+^+(\xi^l_\cC + \xi^r_\cC),
$$
defined by means of equations (\ref{4}) and (\ref{8}).
The reduced orbit
\be
\cO_\red := \phi^{-1}(0)/M
\label{28}\ee
is a stratified symplectic space in general \cite{OR}.  In particular,
$\cO_\red$ contains a dense open subset which is a symplectic manifold and its complement
is the disjunct union of lower dimensional symplectic manifolds. Accordingly, when talking about the
reduced orbit $(\cO_\red, \Omega_\red)$, $\Omega_\red$ actually denotes  a collection
of symplectic forms on the various strata of $\cO_\red$.

The key result for the characterization  of $\check P_\red$ (\ref{24}) is encapsulated by the
following proposition,
whose formulation contains the functions
\be
w(x):= \frac{1}{\sinh(x)}\quad\hbox{and}\quad \chi(x):=\frac{1}{\cosh(x)}.
\label{32}\ee

\medskip
\noindent
{\bf Proposition 1.} \emph{Every $G$-orbit in the constraint surface $\check P_c$ (\ref{25})
possesses representatives of the form $(e^q, J, \xi^l, \xi^r)$,
where $q\in \check \cA$, $p\in \cA$, $\phi(\xi^l,\xi^r)=0$ and $J$ is given by the  formula
\bea
&& J= p - \xi^l
-w(\ad_q) \circ \pi^+_+ (\xi^r_{\cC^\perp} )
- \coth(\ad_q)\circ \pi_+^+ (\xi^l_{\cC^\perp})
\nonumber\\
&& \phantom{XXX}
+ \pi_-^+(\xi^r_{\cC})
+\chi(\ad_q) \circ \pi_-^+ (\xi^r_{\cC^\perp})
- \tanh(\ad_q)\circ \pi_+^- (\xi^l_{\cC^\perp}).
\label{29}\eea
Every element $(e^q, J, \xi^l, \xi^r)$ of the above specified form
belongs to $\check P_c$, and two such  elements belong to the same $G$-orbit if and
only if they are related by the action of the subgroup $M_\diag < Y_+ \times Y^+$,
under which $q$ and $p$ are invariant and the pair $(\xi^l, \xi^r)$ transforms by (\ref{26}).
Consequently, the space of orbits $\check P_\red$ can be identified as
$$
\check P_\red \simeq (\check \cA \times \cA) \times \cO_\red.
$$
This yields the symplectic identification $\check P_\red \simeq T^* \check\cA \times \cO_\red$,
i.e., the reduced (stratified) symplectic form $\omega_\red$ of $\check P_\red$ can be represented as
\be
\omega_\red = d \langle p, dq \rangle + \Omega_\red.
\label{31}\ee
Here, $T^* \check \cA $ is identified with $\check \cA \times \cA = \{(q,p)\}$ and $(\cO_\red, \Omega_\red)$
is
the reduced orbit (\ref{28}).}

\medskip

Proposition 1 is easily proved by solving the moment map constraint
after ``diagonalizing'' $y\in \check Y$ utilizing the generalized Cartan decomposition (\ref{16}).
The expression (\ref{31}) of $\omega_\red$ follows by evaluation of the
original symplectic form $\omega$ on the ``overcomplete set of representatives'' $\{ (e^q, J, \xi^l, \xi^r)\}$
of the $G$-orbits in $\check P_c$.
The operator functions of $\ad_q$ that appear in (\ref{29}) are well-defined since $q\in \check \cA$ is
regular.
Indeed, $\ad_q$ in (\ref{29}) always acts on $\cC^\perp$, where it is invertible\footnote{
For example, the action of $w(\ad_q)$ in (\ref{29}) is defined by expanding
$w(x)$ as $x^{-1}$ plus a power series in $x$, and  then substituting
$\left(\ad_q\vert_{\cC^\perp}\right)^{-1}$ for $x^{-1}$.}.

Now the formula of the reduced ``kinetic energy''
$\cH = \frac{1}{2} \langle J, J \rangle$ is readily calculated.

\medskip
\noindent
{\bf Proposition 2.}
\emph{The reduction of the free Hamiltonian $\cH$ is given by the following $M$-invariant function,
$\cH_\red$, on $T^* \check\cA \times \phi^{-1}(0)$:
\bea
&& 2\cH_\red(q,p, \xi^l,\xi^r) = \langle p, p\rangle
+ \langle \xi^l_\cC, \xi^l_\cC\rangle
+  \langle \pi_-(\xi^r_\cC), \pi_-(\xi^r_\cC)\rangle
\label{33}\\
&&\phantom{XXXX}
 - \langle w^2(\ad_q)\circ\pi^+( \xi^l_{\cC^\perp}), \pi^+(\xi^l_{\cC^\perp})\rangle
 - \langle w^2(\ad_q)\circ\pi_+( \xi^r_{\cC^\perp}), \pi_+(\xi^r_{\cC^\perp})\rangle
\nonumber\\
&&\phantom{XXXX}
+ \langle \chi^2(\ad_q)\circ\pi^-( \xi^l_{\cC^\perp}), \pi^-(\xi^l_{\cC^\perp})\rangle
 + \langle \chi^2(\ad_q)\circ\pi_-( \xi^r_{\cC^\perp}), \pi_-(\xi^r_{\cC^\perp})\rangle
\nonumber\\
&&\phantom{XXXX}
- 2\langle (w^2\chi^{-1})(\ad_q)\circ\pi^+( \xi^l_{\cC^\perp}), \pi_+(\xi^r_{\cC^\perp})\rangle
+ 2\langle (\chi^2 w^{-1})(\ad_q)\circ\pi_-( \xi^r_{\cC^\perp}), \pi^-(\xi^l_{\cC^\perp})\rangle,
\nonumber\eea
where the notations (\ref{32}) and $\chi^{-1}(x):= \cosh(x)$,
$w^{-1}(x):= \sinh(x)$ are applied.
}

In the special case $\gamma=\theta$, studied in \cite{FP},  the formulae simplify considerably.
Indeed, in this case $\pi_+^- = \pi_-^+ =0$, and thus the second line of equation (\ref{29})
and all terms in the last two lines of (\ref{33}) except the one containing $w^2\chi^{-1}$
disappear.
(This term can be recast in a more friendly form by the identity
$(w^2\chi^{-1})(x) = \frac{1}{2} w^2(\frac{x}{2}) - w^2(x)$.)
Although such simplification does not occur in general, we can interpret $\cH_\red$
as a spin Sutherland type Hamiltonian. This means that we view the components of
$q$ as describing the positions of point particles  moving on the line, whose interaction is
governed by hyperbolic functions of $q$ and ``dynamical coupling parameters'' encoded
by the ``spin'' degrees of freedom represented by $\cO_\red$.

\section{A spinless example}

We now recall the special case \cite{AF} whereby the previously described general construction
leads to the $BC_n$ Sutherland system (\ref{1}).
We start by fixing positive integers $1\leq m < n$.
We then prepare the matrices
$$
Q_{n,n}:=\left[
\begin{array}{cc}
0 & \1_n\\
\1_n & 0
\end{array}\right]\in gl(2n,\bC),
\quad
I_{m}:=\operatorname{diag}(\1_m,-\1_{n-m})
\in gl(n,\mathbb{C}),
$$
where $\1_n$ denotes the $n\times n$ unit matrix,  and introduce also
$$
D_m := \operatorname{diag}(I_{m}, I_{m}) =
\operatorname{diag}(\1_m, -\1_{n-m}, \1_{m}, -\1_{n-m})
\in gl(2n,\bC).
$$
We  realize the group $Y:= SU(n,n)$ as
$$
SU(n,n) = \{ y \in SL(2n,\bC)\,\vert\, y^\dagger Q_{n,n} y = Q_{n,n}\},
$$
and define its involutions $\Theta$ and $\Gamma$ by
$$
\Theta(y):= (y^\dagger)^{-1},
\qquad
\Gamma(y):= D_m \Theta(y) D_m,
\qquad
\forall y\in Y.
$$
The fixed point subgroups  $Y_+$ and $Y^+$  turn out to be isomorphic to $S(U(n) \times U(n))$
and  $S( U(m,n-m) \times U(m,n-m))$, respectively.
We choose the maximal Abelian subspace $\cA$  as
\be
\cA:=
\left\{ q:=\left[
\begin{array}{cc}
\mathbf{q} & 0\\
0 & -\mathbf{q}
\end{array}\right]:
\,\,
\mathbf{q}=\diag(q_1,..., q_n),\,\, q_k\in \bR\right\}.
\label{39}\ee
Its centralizer is $\cC = \cA + \cM$ with
$$
\cM \equiv \cC^+_+ =
\left\{ d:=\ri \left[
\begin{array}{cc}
\mathbf{d} & 0\\
0 & \mathbf{d}
\end{array}\right]:
\,\,
\mathbf{d}=\diag(d_1,..., d_n),\,\, d_k\in \bR,\,\, \tr(d) =0\right\}.
$$
In particular, now $\cC_+^- = \cC_-^+ =\{0\}$.
The ``Weyl chamber'' $\check \cA$ can be chosen as those
elements $q\in \cA$ (\ref{39}) whose components satisfy
Eq.~(\ref{2}).

It is important for us that
both $\cY_+$ and $\cY^+$ possess one-dimensional centres,
whose elements can be viewed also as non-trivial one-point coadjoint orbits
of $Y_+$ and $Y^+$.
The centre of $\cY_+$ is generated  by $C^l:= \ri  Q_{n,n}$, and
the centre of $\cY^+$ is spanned  by
$$
C^r:= \ri  \left[\begin{array}{cc}
0 & I_m\\
 I_m & 0
\end{array}\right].
$$
These elements enjoy the property
$$
C^\lambda \in (\cC^\perp)_+^+
\quad\hbox{for}\quad \lambda =l, r.
$$
Taking non-zero real constants $\kappa$ and $x_0$, we choose the coadjoint orbit of $Y_+$ to be
$$
\cO^l\equiv \cO_{\kappa,x_0} :=
\{x_0 C^l +\xi(u) \vert\, u\in \bC^n, \,  u^\dagger u = 2 \kappa n \},
$$
where
\be
\xi(u):=
\frac{1}{2}
\left[
\begin{array}{cc}
X(u) & X(u)\\
X(u) & X(u)
\end{array}\right]
\quad\hbox{with}\quad
X(u) := \ri \left(u u^\dagger - \frac{u^\dagger u}{n} \1_n\right).
\label{44}\ee
It is not difficult to see that the elements $\xi(u)$ in (\ref{44}) constitute a minimal coadjoint orbit
of an $SU(n)$ block of $Y_+\simeq S(U(n) \times U(n))$. The orbit $\cO^r$ of $Y^+$ is chosen to be
$\{ y_0 C^r\}$ with some $y_0\in \bR$, imposing for technical reasons that $(x_0^2 - y_0^2)\neq 0$.

With the above data, we proved that the \emph{full} reduced phase space $P_\red$ (\ref{22})
is given
by the cotangent bundle $T^* \check \cA$, i.e., $\check P_\red = P_\red$.
Moreover, the reduced free Hamiltonian turned out to yield precisely the
$BC_n$ Sutherland Hamiltonian (\ref{1}). The details can be found in  \cite{AF}.

It is an important feature of our example that $\cO^r$ is a one-point coadjoint orbit that belongs
to $(\cC^\perp)_+^+$. Notice that several terms of (\ref{33}), including the unpleasant
last term, disappear for any such orbit.
An even more special feature of the example is that $\cO_\red$ contains
a single element, which means that no spin degrees of freedom are present.
This can be traced back to the well-known fact that the reductions of the minimal coadjoint
orbits of $SU(n)$  by the maximal torus, at zero moment map value, yield one-point spaces.
This fact underlies all derivations of spinless Sutherland type systems
from free geodesic motion that we are aware of, starting from the classical paper \cite{KKS}.

\section{Conclusion}

In this contribution, we described a general class of Hamiltonian reductions
of free motion on a non-compact simple Lie group.
All spin Sutherland type systems that we obtained are expected to yield integrable systems
after taking into account their complete phase spaces provided by
$P_\red$ (\ref{22}). It could be interesting
to investigate the fine details
of these reduced phase spaces and to also investigate their quantization.
Because of their more immediate physical interpretation, the exceptional spinless
members (like the system (\ref{1}))
of the pertinent family of spin Sutherland type systems  deserve closer attention,
and this may motivate one  to ask about the list of all
spinless cases that can occur in the reduction framework.

\bigskip
\medskip
\noindent{\bf Acknowledgements.}
This work was supported in part
by the Hungarian
Scientific Research Fund (OTKA) under the grant K 77400.


\begin{thebibliography}{99}

\bibitem{OPCim}
M.A. Olshanetsky and A.M. Perelomov,
{\it Explicit solutions of some completely integrable systems,}
Lett. Nuovo Cim. {\bf 17},  97-101 (1976).

\bibitem{KKS}
D. Kazhdan, B. Kostant and S. Sternberg, {\it Hamiltonian group
actions and dynamical systems of Calogero type,} Comm. Pure Appl.
Math. {\bf XXXI},  481-507 (1978).

\bibitem{OPRep}
M.A. Olshanetsky and A.M. Perelomov,
{\it Classical integrable finite-dimensional systems related to Lie algebras},
Phys. Rept. {\bf 71}, 313-400 (1981).

\bibitem{Banff}
S.N.M. Ruijsenaars,
{\it Systems of Calogero-Moser type},
pp.~251-352 in:  Proceedings of the 1994 CRM--Banff Summer School `Particles and Fields',
Springer,   1999.

\bibitem{Suth}
B. Sutherland,
Beautiful Models, World Scientific, 2004.

\bibitem{FK}
L. Feh\'er and C. Klim\v c\'\i k,
{\it Poisson-Lie interpretation of trigonometric Ruijsenaars duality},
Commun. Math. Phys. {\bf 301},  55-104 (2011).


\bibitem{AF}
V. Ayadi and L. Feh\'er,
{\it An integrable $BC_n$ Sutherland model with two types of particles},
Journ. Math. Phys {\bf 52}, 103506 (2011).

\bibitem{Ha}
M. Hashizume,
{\it Geometric approach to the completely integrable Hamiltonian systems attached
to the root systems with signature},
Adv. Stud. Pure Math. {\bf 4}, 291-330 (1984).


\bibitem{FP}
L. Feh\'er and B.G. Pusztai,
{\it A class of Calogero type
reductions of free motion on a simple Lie group},
Lett. Math. Phys. \textbf{79}, 263-277 (2007).


\bibitem{KAH}
H. Schlichtkrull,
{\it Harmonic analysis on semisimple symmetric spaces},
pp. 91-225 in:
G. Heckman and H. Schlichtkrull,
Harmonic Analysis and Special Functions on Symmetric Spaces,
Perspectives in Mathematics 16, Academic Press, 1994.

\bibitem{OR}
J.-P. Ortega and T.S. Ratiu, Momentum Maps and Hamiltonian
Reduction, Progress in Mathematics 222, Birkh\"auser, 2004.

\end{thebibliography}
\end{document}